\documentclass[12pt]{article}
\usepackage[english]{babel}
\usepackage{graphicx,color} 
\usepackage{bm}       
\usepackage{amsmath}
\usepackage{amssymb}
\usepackage{caption}
\usepackage[affil-it]{authblk}

\DeclareMathAlphabet{\mathpzc}{OT1}{pzc}{m}{it}
\newcommand{\ket}[1]{ | #1 \rangle}
\newcommand{\bra}[1]{ \langle #1 |}

\begin{document}

\title{The microscopic structure of $\pi NN$, $\pi N\Delta$ and $\pi\Delta\Delta$
vertices in a hybrid constituent quark model}
\author{Ju-Hyun Jung and Wolfgang Schweiger}
\affil{Institute of Physics, University of Graz, A-8010 Graz,
Austria}
\date{ }
\maketitle

\begin{abstract}
We present a microscopic description of the strong  $\pi NN$, $\pi N\Delta$ and $\pi\Delta\Delta$ vertices. Our starting point is a constituent-quark model supplemented by an additional $3q\pi$ non-valence component. In the spirit of chiral constituent-quark models, quarks are allowed to emit and reabsorb a pion. This multichannel system is treated in a relativistically invariant way within the framework of point-form quantum mechanics. Starting with a common $SU(6)$ spin-flavor-symmetric wave function for $N$ and $\Delta$, we calculate the strength of the $\pi NN$, $\pi N\Delta$ and $\pi\Delta\Delta$ couplings and the corresponding vertex form factors. Our results are in accordance with phenomenological fits of these quantities that have been obtained within purely hadronic multichannel models for baryon resonances.
\end{abstract}

\section{Introduction}
 One of the big deficiencies of conventional constituent-quark models is the fact that all states come out as stable bound states. In nature, however, excited states are rather resonances with a finite decay width. In order to remedy this situation, we study a constituent-quark model with explicit pionic degrees of freedom.  The underlying physics is that of \lq\lq chiral constituent-quark models\rq\rq. This means that the spontaneous chiral-symmetry breaking of QCD produces pions as the associated Goldstone bosons and constituent quarks as effective particles~\cite{Glozman:1995fu}, with the pions coupling directly to the constituent quarks. The occurrence of pions affects then the masses and the structure of the hadrons and leads to resonance-like behavior of hadron excitations. If one assumes instantaneous confinement between the quarks, only \lq\lq bare\rq\rq\ hadrons, i.e. eigenstates of the pure confinement problem, can propagate. As a consequence, pionic effects on hadron masses and structure can be formulated  as a purely hadronic problem with the hadron substructure entering pion-hadron vertex form factors\footnote{Strictly speaking these are vertex form factors of the bare hadrons.}.  In the present contribution we will present predictions for $\pi NN$, $\pi N\Delta$ and $\pi\Delta\Delta$ couplings and vertex form factors, given the $\pi q q$ coupling and an $SU(6)$ spin-flavor symmetric model for the $3q$ wave function of the nucleon and the $\Delta$.

\section{Formalism}
\begin{figure}[t!]
\begin{center}
\includegraphics[width=0.98\textwidth]{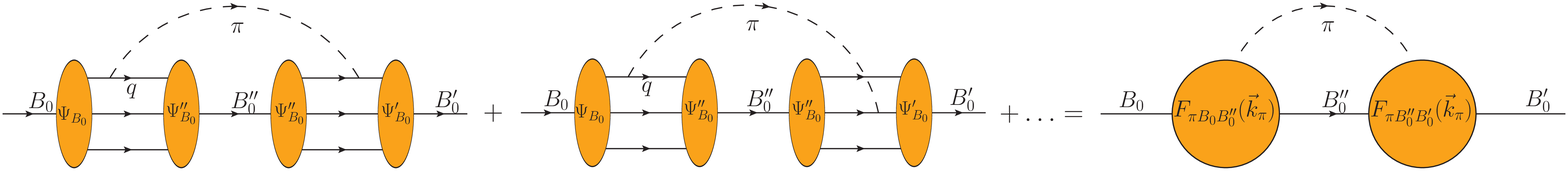}
\end{center}
\caption{Graphical representation of the kernel
$\bra{B_0^\prime}\hat{V}^{\mathrm{opt}}_\pi(m)\ket{B_0}$ needed to solve the mass-eigenvalue equation~(\ref{eq:Mphys}).}
\label{fig:piloop}\end{figure}

Our starting point for calculating the strong $\pi NN$, $\pi N\Delta$ and $\pi\Delta\Delta$ couplings and form factors is the mass-eigenvalue problem for 3 quarks that are confined by an instantaneous potential and can emit and reabsorb a pion. To describe this system in a relativistically invariant way, we make use of the point-form of relativistic quantum mechanics. Employing the Bakamjian-Thomas construction, the overall 4-momentum operator $\hat{P}^\mu$ can be separated into a free 4-velocity operator $\hat{V}^\mu$ and an invariant mass operator $\hat{\mathcal{M}}$ that contains all the internal motion, i.e. $\hat{P}^\mu=\hat{\mathcal{M}}\, \hat{V}^\mu$~\cite{Biernat:2010tp}. Bakamjian-Thomas-type mass operators are most conveniently represented by means of velocity states $\vert V;  {\bf k}_1, \mu_1; {\bf k}_2, \mu_2; \dots ; {\bf k}_n, \mu_n\rangle$, which specify the system by its overall velocity $V$ ($V_\mu V^\mu=1$), the CM momenta ${\bf k}_i$ of the individual particles and their (canonical) spin projections $\mu_i$~\cite{Biernat:2010tp}. Since the physical baryons of our model contain, in addition to the $3q$-component, also a $3q\pi$-component, the mass eigenvalue problem can be formulated as a 2-channel problem of the form
\begin{equation}
\left(
\begin{array}{cc} \hat{M}^{\mathrm{conf}}_{3q}&\hat{K}_\pi\\\hat{K}_\pi^\dag& \hat{M}^{\mathrm{conf}}_{3q\pi}
\end{array}\right)
\left(\begin{array}{c}\ket{\psi_{3q}}\\\ket{\psi_{3q\pi}}
\end{array}\right)=m
\left(\begin{array}{c}\ket{\psi_{3q}} \\ \ket{\psi_{3q\pi}}
\end{array}\right)\, ,
\end{equation}
with $\ket{\psi_{3q}}$ and $\ket{\psi_{3q\pi}}$ denoting the two Fock-components of the physical baryon states $\ket{B}$. The mass operators on the diagonal contain, in addition to the relativistic particle energies, an instantaneous confinement potential between the quarks. The vertex operator $\hat{K}_\pi^{(\dag)}$ connects the two channels and describes the absorption (emission) of the $\pi$ by one of the quarks. Its velocity-state representation can be directly connected to a corresponding field-theoretical interaction Lagrangean~\cite{Biernat:2010tp}. We use a pseudovector interaction Lagrangean for the $\pi q q$-coupling
\begin{equation}\label{eq:piqqvertex}
\mathcal{L}_{\pi q q}(x) = -\frac{f_{\pi q q}}{m_\pi} \left(\bar{\psi}_{q}(x)\gamma_{\mu}\gamma_{5}\vec{\tau}\psi_{q}(x)\right)\cdot\partial^{\mu}\vec\phi_{\pi}(x),
\end{equation}
where the \lq\lq $\cdot$\rq\rq -product has to be understood as product in isospin space. After elimination of the $3q\pi$-channel the mass-eigenvalue equation takes on the form
\begin{equation}\label{eq:Mphys}
\bigl[\hat{M}_{3q}^{\mathrm{conf}} +\underbrace{\hat{K}_\pi(m-\hat{M}_{3q\pi}^{\mathrm{conf}})^{-1} \hat{K}_\pi^\dag}_{\hat{V}^{\mathrm{opt}}_\pi(m)}\bigr] \ket{\psi_{3q}} = m \, \ket{\psi_{3q}} \, ,
\end{equation}
where $\hat{V}^{\mathrm{opt}}_\pi(m)$ is an optical potential that describes the emission and reabsorption of the pion by the quarks. One can now solve Eq.~(\ref{eq:Mphys}) by expanding the ($3q$-components of the) eigenstates in terms of eigenstates of the pure confinement problem, i.e. $\ket{\psi_{3q}}=\sum_{B_0} \alpha_{B_0}\,\ket{B_0}$, and determining the open coefficients $\alpha_{B_0}$. Since the particles which propagate within the pion loop are also bare baryons (rather than quarks), the problem of solving the mass eigenvalue equation~(\ref{eq:Mphys}) reduces then to a pure hadronic problem, in which the dressing and mixing of bare baryons by means of pion loops produces finally the physical baryons (see Fig.~\ref{fig:piloop}). As also indicated in Fig.~\ref{fig:piloop}, the quark substructure determines just the coupling strengths at the pion-baryon vertices and leads to vertex form factors.
To set up the mass-eigenvalue equation on the hadronic level one needs matrix elements $\bra{B_0^\prime}\hat{V}^{\mathrm{opt}}_\pi(m)\ket{B_0}$ of the optical potential between bare baryon (velocity\footnote{We suppress this velocity dependence since it factors out and has no influence on the mass spectrum.}) states. The general structure of these matrix elements is ($B_0$ and $B_0^{\prime}$ are at rest)
\begin{equation}\label{eq:vopt}
\bra{B_0^\prime}\hat{V}^{\mathrm{opt}}_\pi(m)\ket{B_0} \propto \sum_{B_0^{\prime\prime}}\int \frac{d^3 k_\pi^{\prime\prime}}{2 \sqrt{m_\pi^2+{\vec{k}_\pi^{\prime\prime\,^2}}}}\, J^{5\ast}_{\pi B_0^{\prime\prime} B_0^{\prime}}(\vec{k}_\pi^{\prime\prime})\, \frac{1}{m-m_{B_0^{\prime\prime}\pi}} \, J^{5}_{\pi B_0^{\prime\prime} B_0}(\vec{k}_\pi^{\prime\prime})\ ,
\end{equation}
where $m_{B_0^{\prime\prime}\pi}$ is the invariant mass of the $B_0^{\prime\prime}\pi$ system in the intermediate state and spin- as well as isospin dependencies have been suppressed.

For the cases we are interested in, i.e. the $N$ and the $\Delta$, the currents occurring in Eq.~(\ref{eq:vopt}) can be cast into the form\footnote{Note that this form exhibits the correct chiral properties and avoids problems with superfluous spin degrees of freedom when treating spin-3/2 fields covariantly by means of Rarita-Schwinger spinors~\cite{Pascalutsa:2005nd}.}:
\begin{eqnarray}\label{eq:currents}
J^{5}_{\pi N_0 N_0}(\vec{k}_\pi) & = & i\, \frac{f_{\pi N_0 N_0}}{m_\pi} F_{\pi N_0 N_0}(\vec{k}_\pi^2)\, \bar{u}(-\vec{k}_\pi)\gamma_{\mu}\gamma_{5}u(\vec{0})\, k_{\pi}^{\mu}\, ,\nonumber\\
J^{5}_{\pi \Delta_0 \Delta_0}(\vec{k}_\pi) & = &\frac{f_{\pi \Delta_0 \Delta_0}}{m_\pi m_{\Delta_0}} F_{\pi \Delta_0 \Delta_0}(\vec{k}_\pi^2)\, \epsilon^{\mu\nu\rho\sigma}\,\bar{u}_{\mu}(-\vec{k}_\pi)\,u_{\nu}(\vec{0})\,  k_{\Delta_0,\rho}\, k_{\pi,\sigma} \, ,\nonumber\\
J^{5}_{\pi N_0 \Delta_0}(\vec{k}_\pi) & = & -i \frac{f_{\pi N_0 \Delta_0}}{m_\pi m_{\Delta_0}}\, F_{\pi N_0 \Delta_0}(\vec{k}_\pi^2)\, \epsilon^{\mu\nu\rho\sigma}\, \bar{u}(-\vec{k}_\pi)\gamma_{\sigma}\gamma_{5} u_{\nu}(\vec{0})\, k_{\Delta_0,\mu}\,k_{\pi,\rho}\, ,\nonumber \\
J^{5}_{\pi \Delta_0 N_0}(\vec{k}_\pi) & = & i \frac{f_{\pi N_0 \Delta_0}}{m_\pi m_{\Delta_0}}\, F_{\pi \Delta_0 N_0}(\vec{k}_\pi^2)\, \epsilon^{\mu\nu\rho\sigma}\, \bar{u}_\nu(-\vec{k}_\pi)\gamma_{5}\gamma_{\sigma} u(\vec{0})\, k_{\Delta_0,\mu}\,k_{\pi,\rho}\, ,
\end{eqnarray}
where $u(.)$ is the Dirac spinor of the nucleon and $u_\mu(.)$ the Rarita-Schwinger spinor of the $\Delta$. Here we have again suppressed the isospin dependence and also omitted the spin labels. From Eqs.~(\ref{eq:vopt}) and (\ref{eq:currents}) one can then infer the analytical expression for the  combination $f_{\pi B_0^\prime B_0}\, F_{\pi B_0^\prime B_0}(\vec{k}_\pi^2)$ in terms of quark degrees of freedom. It is an integral over the (independent) quark momenta involving the $3q$ wave function of the in- and outgoing (bare) baryons, the pseudovector quark current as resulting from the Lagrangean (\ref{eq:piqqvertex}) and some kinematical as well as Wigner-rotation factors~\cite{Kupelwieser:2016}.

Assuming a scalar isoscalar confinement potential, the masses of the bare nucleon and the bare $\Delta$ are degenerate, the momentum part of the wave function will be the same and the spin-flavor part of the wave function is $SU(6)$ symmetric. Rather than solving the confinement problem for a particular potential, we thus parameterize the momentum part of the $3q$ wave function of $N_0$ and $\Delta_0$ by means of a Gaussian
\begin{equation}
\psi_{3q}^{N_0,\Delta_0}(\vec{k}_{q_1},\vec{k}_{q_2},\vec{k}_{q_3})\propto \exp\left(-\alpha^2 (\vec{k}_{q_1}^2+\vec{k}_{q_2}^2+\vec{k}_{q_3}^2) \right)\, ,\quad \vec{k}_{q_1}+\vec{k}_{q_2}+\vec{k}_{q_3}=\vec{0}\, ,
\end{equation}
and choose an appropriate value for the mass of $N_0$ and $\Delta_0$, i.e. $M_{N_0}=M_{\Delta_0}=:M_0$. The parameters of our model are therefore the oscillator parameter $\alpha$, the $N_0$ and $\Delta_0$ mass $M_0$, the constituent-quark mass $m_q:=m_u=m_d$ and $f_{\pi q q}$, the $\pi q q$ coupling strength. For fixed $m_q=263$~MeV we  have adapted the remaining parameters such that the physical $N$ and $\Delta$ masses, resulting from the mass renormalization due to pion loops (with $N_0$ and $\Delta_0$ intermediate states), agree with their experimental values. This gives us for the remaining parameters $M_0=1.552$~GeV, $\alpha=2.56$~GeV$^{-1}$ and $f_{\pi q q}=0.6953$.

\section{Results and Outlook}
\begin{figure}[t!]
\includegraphics[scale=0.4]{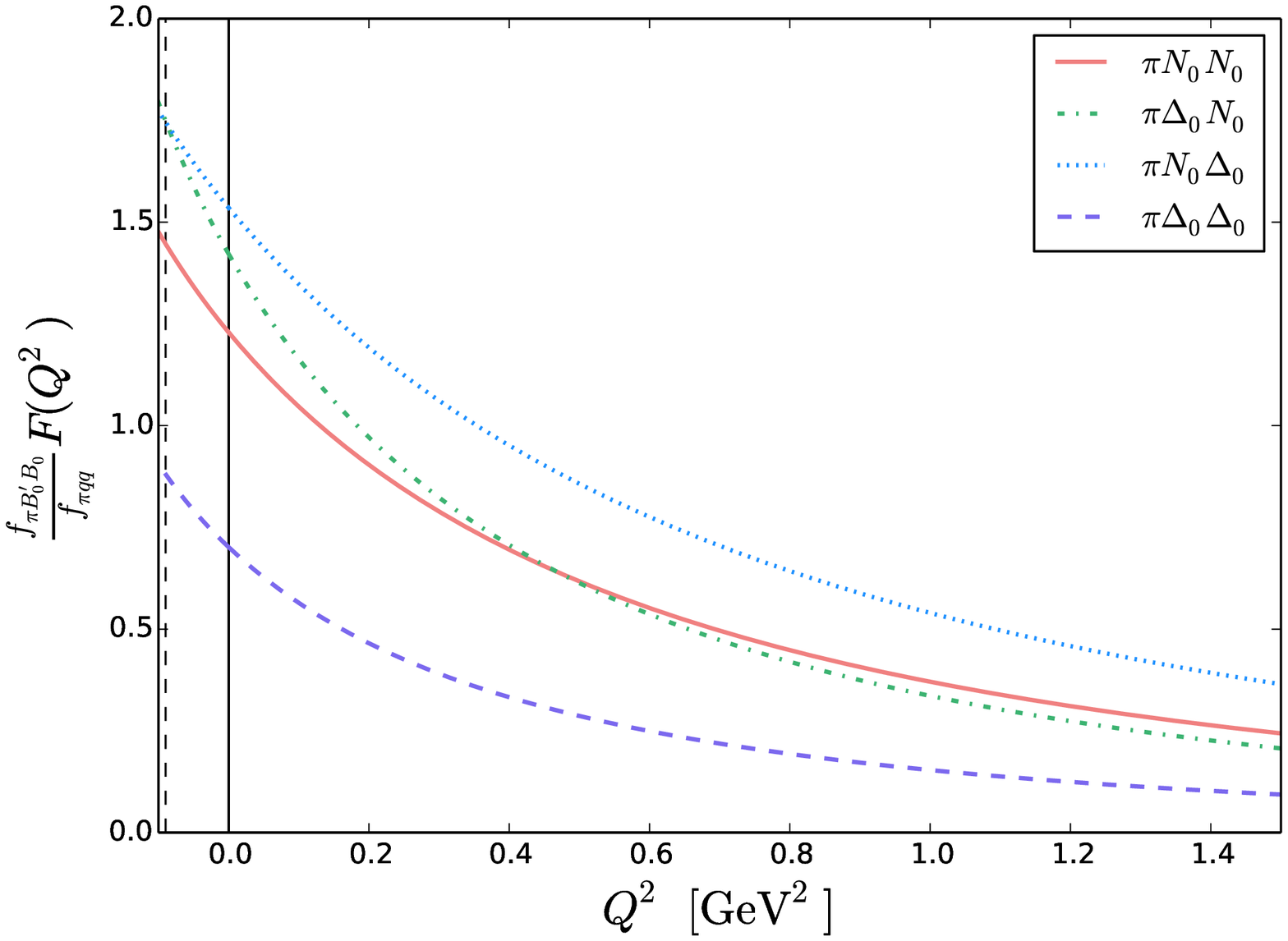}\includegraphics[scale=0.4]{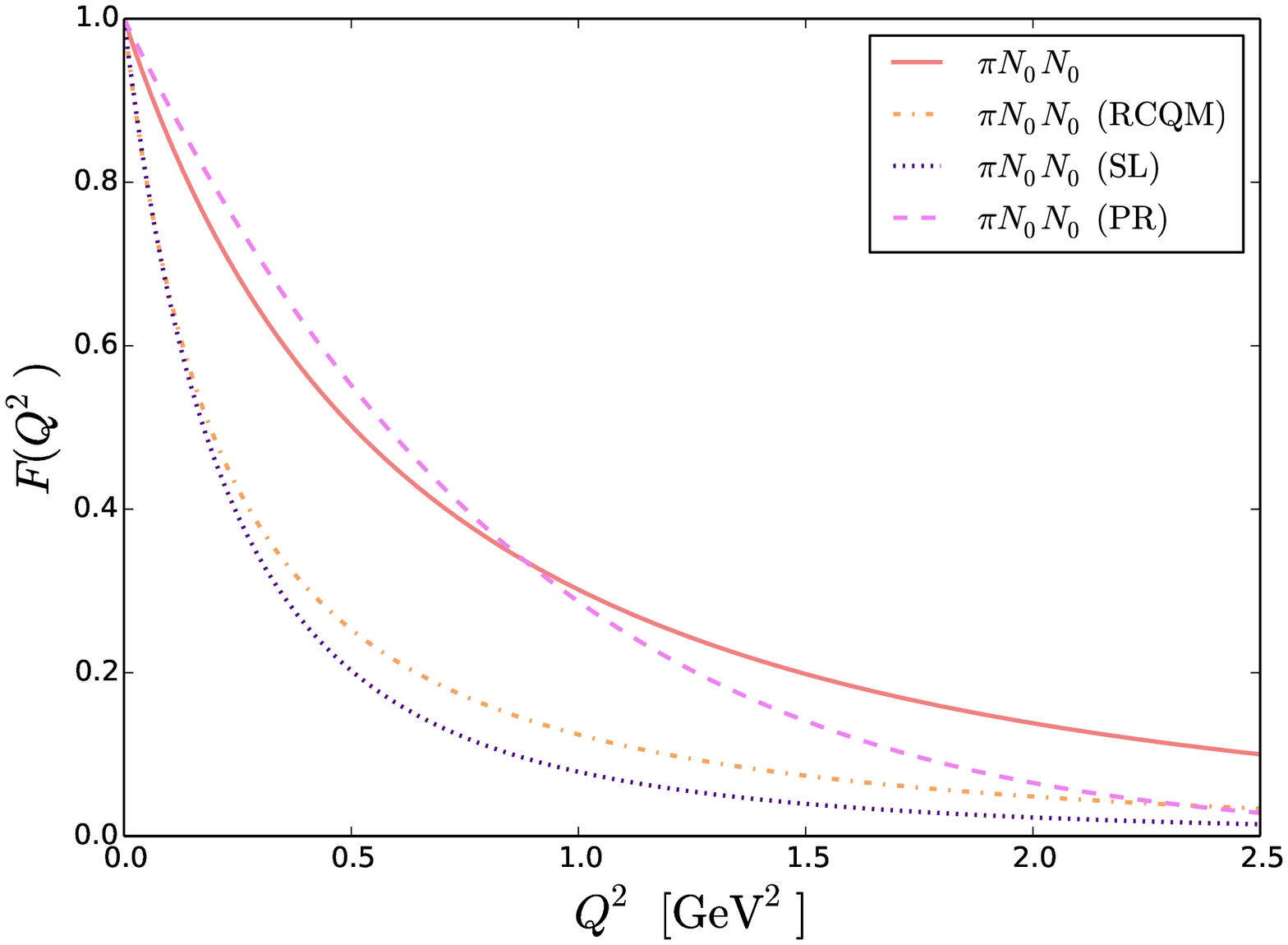}
\caption{The left plot shows the (unnormalized) $\pi N_0 N_0$, $\pi \Delta_0 \Delta_0$, $\pi N_0 \Delta_0$, and $\pi \Delta_0 N_0$ form factors as functions of $Q^2=-2 M_0 (M_0- (M_0^2+\vec{k}_\pi^2)^{1/2})$. In the right plot the $Q^2$ behavior of $F_{\pi N_0 N_0}$ (normalized to $1$ at $Q^2=0$) is compared to the outcome of another relativistic constituent-quark model (RCQM)~\cite{Melde:2008dg} and of phenomenological fits obtained within two purely hadronic dynamical coupled-channel models~\cite{Kamano:2013iva,Polinder:2005sn} (SL and PR).}\label{fig:ffs}
\end{figure}
Having fixed the parameters of our model, we are now able to make predictions for the strong $\pi N_0 N_0$, $\pi \Delta_0 \Delta_0$, $\pi N_0 \Delta_0$, and $\pi \Delta_0 N_0$ couplings and form factors. The left plot of Fig.~\ref{fig:ffs} shows these (unnormalized) form factors as function of the (negative) four-momentum transfer squared (analytically continued to small time-like momentum transfers). It is worth noting that $F_{\pi \Delta_0 N_0}$ and $F_{\pi N_0 \Delta_0}$ do not agree. This is, of course, no surprise, since in the first case the $N_0$ is real and the $\Delta_0$ virtual, whereas it is just the other way round in the second case. The form factors describe thus completely different kinematical situations, but they coincide at a particular negative (i.e. unphysical) value of $Q^2$. Since there is only one coupling strength at the $\pi N_0\Delta_0$-vertex (i.e. $f_{\pi \Delta_0 N_0}=f_{\pi N_0\Delta_0}$, see Eq.~(\ref{eq:currents})), this is the natural point to normalize the form factors and extract the coupling constants. Its value $Q_0^2=-0.090$~GeV$^2$ is close to the standard normalization point, namely the pion pole $Q_0^2=-m_\pi^2$. Comparing the resulting coupling strengths, we get the ratio $f_{\pi N_0\Delta_0}:f_{\pi N_0 N_0}:f_{\pi \Delta_0\Delta_0}=1.208:1:0.608$. This should be compared with the prediction from the non-relativistic constituent-quark model assuming $SU(6)$ spin-flavor symmetry, i.e. $f_{\pi N\Delta}:f_{\pi N N}:f_{\pi \Delta\Delta}=4 \sqrt{2}/5:1:4/5=1.13:1:0.8$~\cite{Brown:1975di}. The differences can solely be ascribed to relativistic effects and are obviously significant, in particular for the $\pi \Delta_0\Delta_0$-vertex. Remarkably, our results resemble very much those needed in dynamical coupled-channel models, e.g. $f_{\pi N\Delta}:f_{\pi N N}:f_{\pi \Delta \Delta}=1.26:1:0.42$ in Ref.~\cite{Kamano:2013iva}.

In the right plot of Fig.~\ref{fig:ffs} our result for $F_{\pi N_0 N_0}$ is compared with the outcome of another relativistic constituent-quark model~\cite{Melde:2008dg} and with two parameterizations of this form factor that have been used in dynamical coupled-channel models~\cite{Polinder:2005sn,Kamano:2013iva}. Up to $Q^2\approx 1$~GeV$^2$ our prediction is comparable with the form factor parametrization of Ref.~\cite{Polinder:2005sn}, but for higher $Q^2$ it falls off slower. The form factors of Refs.~\cite{Melde:2008dg,Kamano:2013iva} fall off much faster already at small $Q^2$. Deviations of our result from the one of Ref.~\cite{Melde:2008dg} have their origin in different $3q$ wave functions of the nucleon, but also in different kinematical and spin-rotation factors entering the microscopic expression for the pseudovector current of the nucleon.

Having determined the $\pi N_0 N_0$, $\pi \Delta_0 \Delta_0$ and $\pi N_0 \Delta_0$ vertices from a microscopic model, we are now in the position to calculate the electromagnetic form factors of physical nucleons and Deltas and determine the effect of pions on their electromagnetic structure. First exploratory calculations for the nucleon show that visible effects can be expected for $Q^2\lesssim 0.5$~GeV$^2$~\cite{Kupelwieser:2016}. It will, of course, be more interesting to investigate electromagnetic $\Delta$ and $N\rightarrow \Delta$ transition form factors, where pionic effect are expected to play a more significant role.

\end{document}